# Long cavity spectral disperser at sub-picometer resolution. Design and analysis

François Hénault[1], Yan Feng[1,2]

[1] Optical Engineering Consulting, 740 Chemin d'Argevillières, 07000 Veyras – France

[2] LIRA, Observatoire de Paris, Université PSL, Université Paris Cité, Sorbonne Université, CY Cergy Paris Université, CNRS, 75014 Paris – France

## ABSTRACT

In many applications of spectrometry, a very high spectral resolution is of paramount importance for technologies such as wavelength division multiplexing, femtosecond laser pulse shaping, chemical analysis of gases, or astrophysics observations. A few techniques achieving such goal already exist as listed in the introducing paragraphs. In this article is described a long cavity spectral disperser that is likely to be integrated into a spectrometer and having the potential to attaining unsurpassed spectral resolving power equal to $\lambda/d\lambda$ = ten millions or more. The basic relations of its angular dispersion, free spectral range, resolving power, transmission and contrast are established and a preliminary optimized design is presented

**Keywords** Instrumentation, Spectroscopy, Spectral resolution, Dispersive optics and components, Intererometric cavity

## 1. INTRODUCTION

Highly dispersive optical components or systems are of uttermost importance in the fields of Wavelength division multiplexing (WDM) and light wave communications (Shirasaki, 1996, Okamoto, 1999), femtosecond laser pulse shaping (Weiner, 2000), chemical analysis of atmospheric gases for environment monitoring (Adler et al., 2010, Nugent-Glandorf et al., 2012), and many others such as biomedicine, communication networks, chemistry and metallurgy. In the field of astronomy, they are very popular in view of analyzing the spectra of celestial objects (Hall et al., 1994). Currently, they are used for identifying the presence of extrasolar planets orbiting around their parent star (Donati et al., 2020) and characterizing their atmospheres. Different designs of high resolution spectrometers already exist and are available on the market, such as the scanning Fabry-Pérot interferometer (Paulo Carmo et al., 2012, Kuhn et al., 2021) or the Fourier transform spectrometer (Horlick, 1968, Lindner et al., 2022). Very high spectral resolutions can also be achieved with static optical components, such as echelle gratings (Zhang et al., 2022) and the Virtually imaged phased array (VIPA) that may provide higher performance and requires a smaller volume (Vega, Weiner & Lin; 2003, Xiao, Weiner & Lin, 2004). For now, all the cited devices or equipments routinely provide relative spectral resolutions in the range $[10^{+5} – 10^{+6}]$. However, some of the previous applications such as gases chemical analysis or outstandingly demanding astronomic goals may require significantly higher numbers. In this article is described a long Cavity spectral disperser (CSD) based on an interferometric component that may achieve higher spectral resolution. Here are presented the optical layout of the CSD, a theoretical analysis enabling to evaluate its potential performance, and a short conclusion.

## 2. OPTICAL DESIGN

Tie CSD component can be seen as a variant of the Franson interferometer (Franson, 1989) originally intended to solve the famous Einstein-Podolsky-Rosen (EPR) paradox. The apparatus is depicted in Figure 1 and the employed parameters, scientific notations, and numerical values are listed in Table 1. It starts from a parallel input beam at the bottom left corner of the figure and directed along the X-axis, whose dimensions are $p \times W$ along the Y and Z axes respectively, which can be achieved by using anamorphic fore-optics not shown on the figure. Impinging the first beamsplitter BS1 at point O, the rays can either be transmitted along the X-axis and exit the cavity, or reflected along the Y-axis. Those rays then encounter two Folding mirrors (FM1 and FM2) successively and finally reach the second beamsplitter BS2 which is slightly shifted by a quantity equal to $p$ in the +Y direction. Rays reflected by BS2 are thus separated laterally from the primary beam by the same distance. Returning back to BS1 they can either exit the cavity or start a second cavity loop, and this process can be repeated a certain number of times $M$. Assuming that $M = W / p$, the full beam exiting the CSD on the +X side now appears as a series of adjacent sup-pupils of dimensions are $p \times W$, having experienced varying Optical path differences (OPD), and forming together a squared output pupil of $W$-side. One should note that the CSD also produces a second exit beam along the –Y axis, this time being rectangular with dimensions $(W - p) \times W$. The two output composite parallel beams are later focused on detector arrays, not shown in the figure.

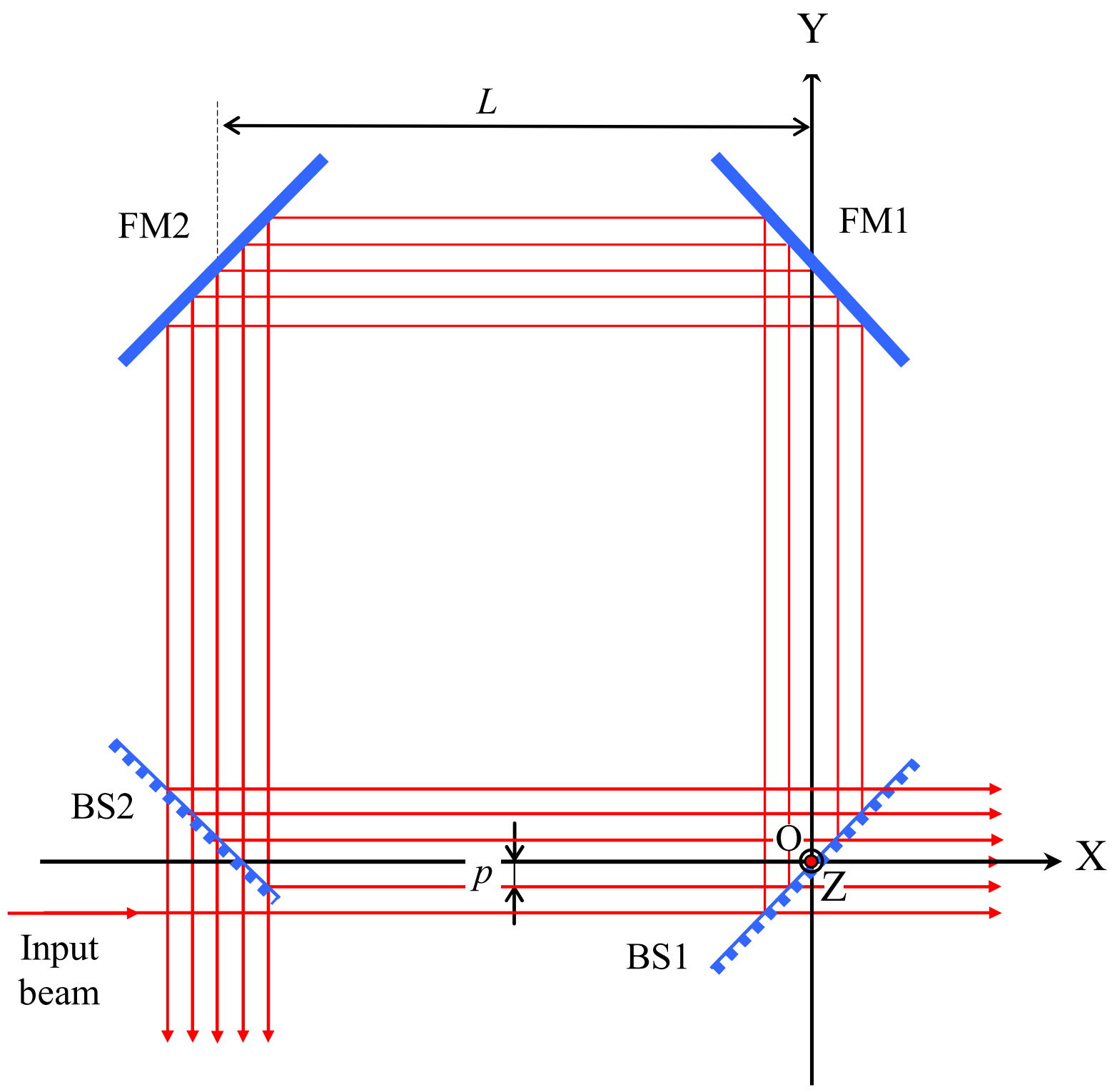

Fig.1: Schematic design of the CSD showing its two output ports along the +X and –Y axes. The Z-axis is directed upward and perpendicular to the plane of the figure.

Table 1: List of employed parameters, scientific notations, and their numerical values.

| Parameter | Symbol | Value |
|---|---|---|
| Wavelength | $\lambda$ | 1 µm |
| Incidence angle | $i$ | 0 deg |
| Diffracted angle | $\beta$ | +30 deg |
| Cavity side | $L$ | 100 mm |
| Cavity length | $\Delta_0 = 4\,L$ | 400 mm |
| Exit pupil width | $W$ | 20 mm |
| Sub-pupils number along Y-axis | $M$ | 20 |
| Sub-pupils width along Y-axis | $p = W / M$ | 1 mm |

## 3. ANALYSIS

The Figure 2 illustrates the calculation of the OPD generated between two adjacent sub-output beams as function of the main parameters defined in Table 1. By definition it is equal to:

$$\delta_0 = [OABCD] - |OD| \sin(\pi/4 + \beta)\,; \tag{1}$$

where $[OABCD]$ is the total path of a ray impinging point O under an incidence angle $i$ and crossing the whole cavity, $|OD|$ is the distance between points O and D, and $\beta$ is the diffraction angle. The analytic form of $\delta_0$ is determined by following the optical ray reflected by BS1, FM1, FM2 and BS2 successively (see Figure 2). Although a bit long and tedious, this calculation is not difficult and leads to the following expression:

$$\delta_0 = \frac{\Delta_0}{\cos i - \sin i} - |OD|\sin(\pi/4 + \beta) \quad \text{with} \tag{2}$$

$$|OD| = \sqrt{2p^2 + 4\sin i \frac{\Delta_0}{\cos i - \sin i}\left(p + \frac{\sin i}{2}\frac{\Delta_0}{\cos i - \sin i}\right)}. \tag{3}$$

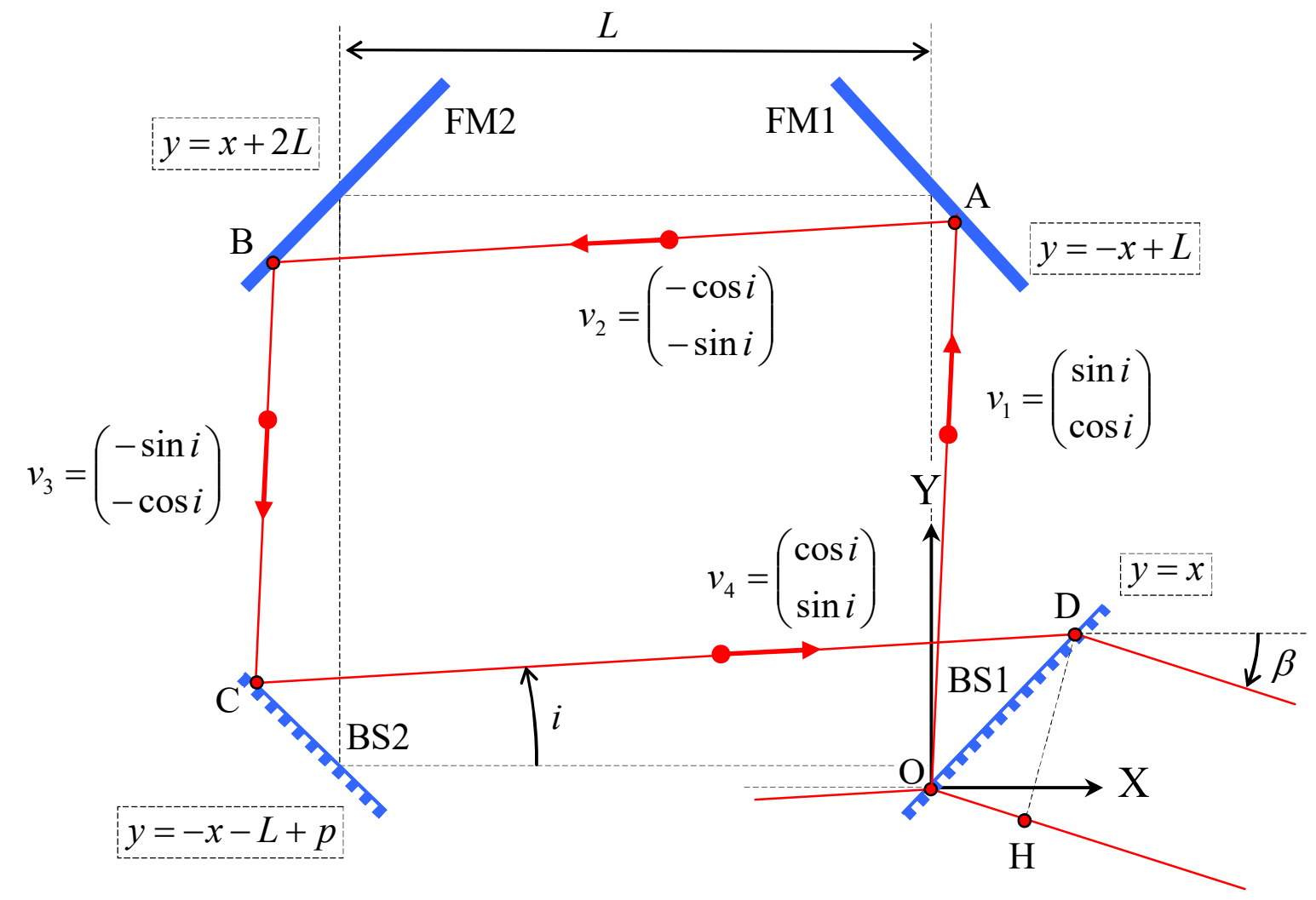

Fig. 2. Computation the OPD generated between two adjacent sub-beams. The coordinate relations in the planes of the beamsplitters and of the fold mirrors are indicated by dashed line boxes.

From Eq. 2 the classical equation of diffraction gratings writes as:

$$\delta_0 = \frac{\Delta_0}{\cos i - \sin i} - |OD|\sin(\pi/4 + \beta) = n\lambda \tag{4}$$

where $n$ is the diffraction order. The angular dispersion $\partial\beta / \partial\lambda$ of the CSD is then found to be:

$$\frac{\partial\beta}{\partial\lambda} = \frac{1}{\lambda}\left[\frac{\Delta_0}{|OD|(\cos i - \sin i)\cos(\pi/4 + \beta)} - \tan(\pi/4 + \beta)\right]. \tag{5}$$

A sensitivity analysis of the angular dispersion with respect to the different parameters such as the incidence angle $i$, the cavity length $\Delta_0$, the number of output sub-pupils $M = W / p$, and the diffraction angle $\beta$ has been carried out. It showed that $\partial\beta / \partial\lambda$ rapidly drops down when the incidence angle $i$ differs from zero, thus this value will be kept in the remainder of the study. The Figure 3 displays the sensitivity curves of $\partial\beta / \partial\lambda$ as a function of the diffraction angle $\beta$ for different values of the $\Delta_0$ and $M$ parameters. It can be seen that the angular dispersion is roughly proportional to the both parameters, or equivalently it is inversely proportional to the width $p$ of the output sub-pupils along the Y-axis.

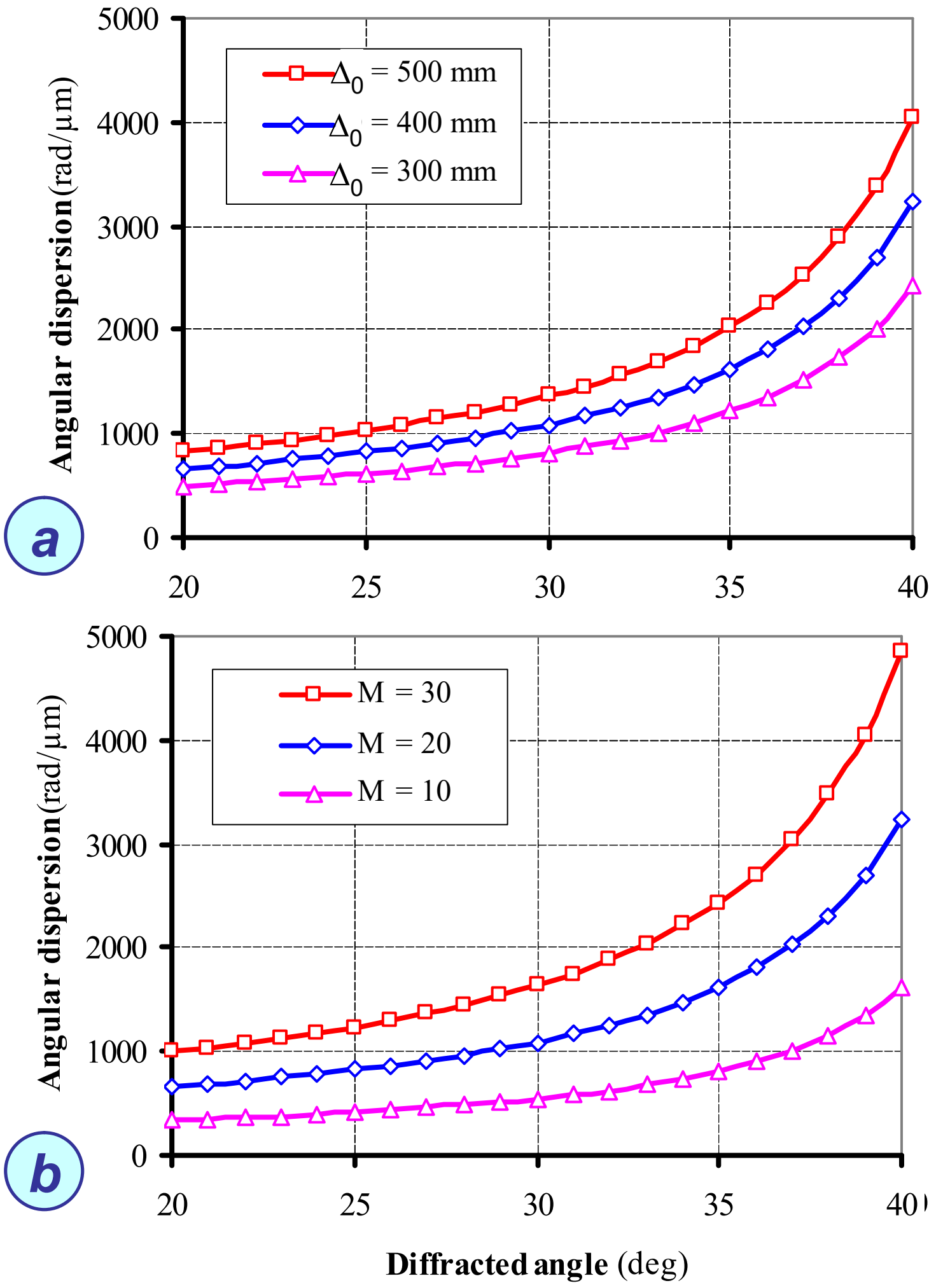

Fig.3: Sensitivity curves of the CSD angular dispersion as function of the diffraction angle for different values of the cavity length $\Delta_0$ (a) and of the number $M$ of cavity loops (*b*).

Perhaps more important than the angular dispersion $\partial\beta / \partial\lambda$ is the spectral resolving power $R_\lambda = \lambda / d\lambda$ of the CSD. The latter is estimated following two steps: firstly the relative Free spectral range (FSR) $\Delta\lambda / \lambda$ is deduced from Eq. 2 as the difference between two adjacent diffraction orders $n$, thus leading to:

$$\frac{\Delta\lambda}{\lambda} = \frac{\lambda}{\Delta_0(\cos i - \sin i) - |OD|\sin(\pi/4 + \beta)} \tag{6}$$

Secondly, the resolving power has to be estimated in the image plane of the CSD, where the generated interferometric fringe pattern is equal to the square modulus of the complex amplitude formed on the detector array. For the output port along X-axis, the latter is equal to:

$$A_X(\lambda, \beta) = \sqrt{T}\sum_{m=0}^{M-1} R^m \exp(2i\pi m[\delta_0 + p(i+\beta)]/\lambda), \tag{7}$$

where $R$ is the intensity reflectance coefficient of both beamsplitters BS1 and BS2, $T = 1 - R$, and the reflectance of the fold mirrors is assumed to be unity to alleviate the formulas. Hence, assuming that there is no limitation due to the pixel size of the detector and using the condensed notation $\xi(\lambda,\beta) = 2\pi[\delta_0+p(i+\beta)] / \lambda$, the intensity of the fringe pattern can be written as:

$$I_X(\lambda,\beta) = T\frac{1+R^{2M} - 2R^M \cos M\xi(\lambda,\beta)}{1+R^2 - 2R\cos\xi(\lambda,\beta)}. \tag{8a}$$

Similarly the intensity generated from the –Y output port is equal to:

$$I_Y(\lambda,\beta) = RT\frac{1+R^{2(M-1)} - 2R^{M-1} \cos(M-1)\xi(\lambda,\beta)}{1+R^2 - 2R\cos\xi(\lambda,\beta)} \tag{8b}$$

The structure of the spectral response $I_X(\lambda,\beta)$ in Eq. 8a as function of the wavelength $\lambda$ is illustrated in Figure 4. It allows to defining the smallest discernable wavelength difference element $d\lambda$. Assuming it to be located at the half maximum of the main peak response, it is linked to the FSR by the relation $d\lambda = \Delta\lambda / 2M$. Then the spectral resolving power of the CSD is finally equal to:

$$R_\lambda = \frac{2M}{\lambda}\left(\frac{\Delta_0}{\cos i - \sin i} - |OD|\sin(\pi/4+\beta)\right) \tag{9}$$

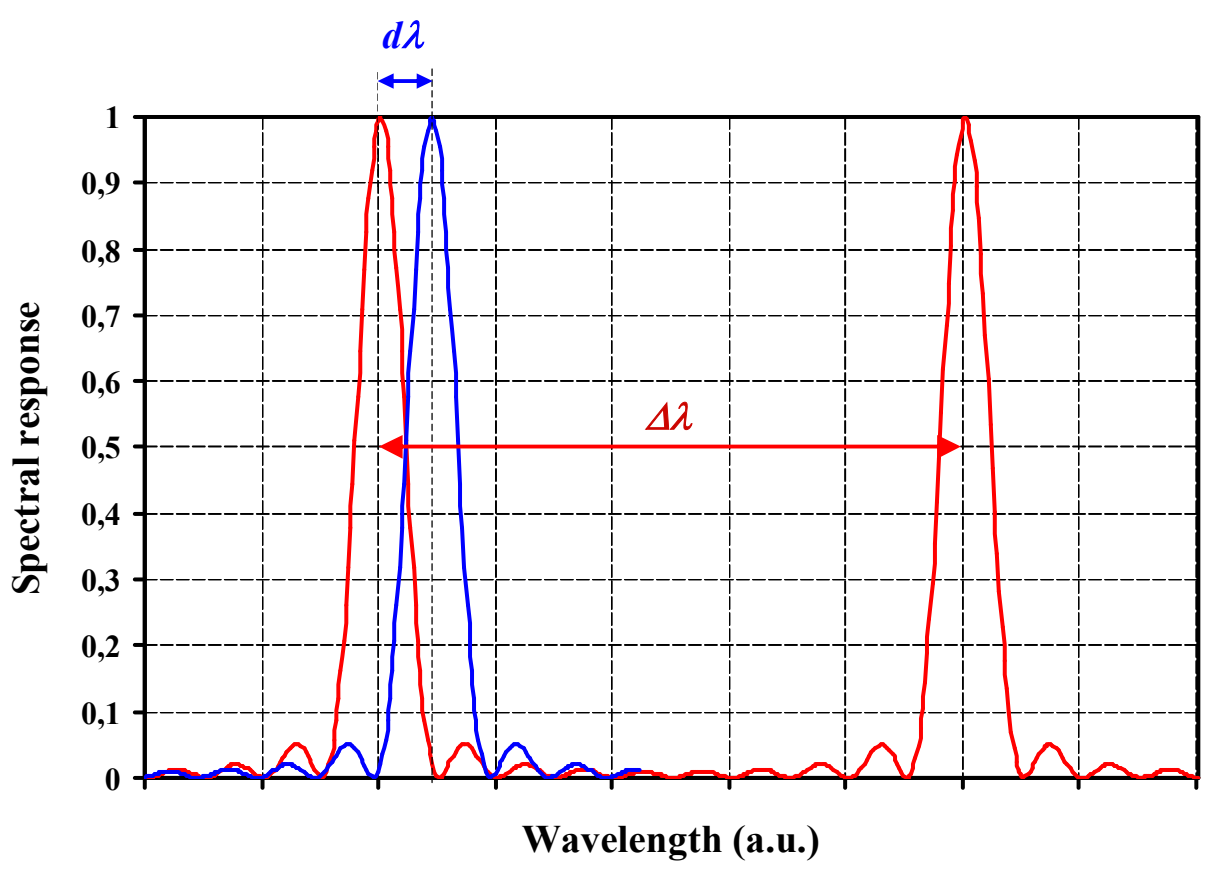

Fig.4: Defining the smallest wavelength difference element $d\lambda$.

From the relations 8 the global average transmissions $I_X$ and $I_Y$ of the output beams at the +X and –Y ports can also be estimated as function of $R$. Neglecting the cosine modulation terms, they can be written as:

$$I_X(R) = T\sum_{m=0}^{M-1}\left(R^2\right)^m = T\frac{1-\left(R^2\right)^M}{1-R^2}, \quad \text{and} \tag{10a}$$

$$I_Y(R) = TR\sum_{m=0}^{M-2}\left(R^2\right)^m = TR\frac{1-\left(R^2\right)^{M-1}}{1-R^2}. \tag{10b}$$

There it can be seen that when the number $M$ of cavity loops tends toward infinity they reduce to $I_X(R) = T/(1-R^2)$ and $I_Y(R) = TR/(1-R^2)$. Therefore $I_X(R) + I_Y(R) = 1$, which satisfies the principle of energy conservation. In a similar way, the contrast function of the fringes dispersed by the CSD can be approximated as:

$$C_X(R) = 2R^M / \left(1 + R^{2M}\right) \tag{11}$$

along the +X axis. Plots of the transmission curves $I_X(R)$, $I_Y(R)$ and of the contrast function $C_X(R)$ as functions of the BS reflectance factor $R$ are shown in Figure 5-a and Figure 5-b respectively, for different values of the cavity loop number $M$. Careful examination of these curves reveals that a good compromise between the transmission and contrast can be found around typical values of $R = 0.97$ and $M = 20$. In that sense the behavior of the CSD is somewhat different from those of the Fabry-Pérot interferometer and of the VIPA, where the reflectance factor $R$ should be as close as possible to unity.

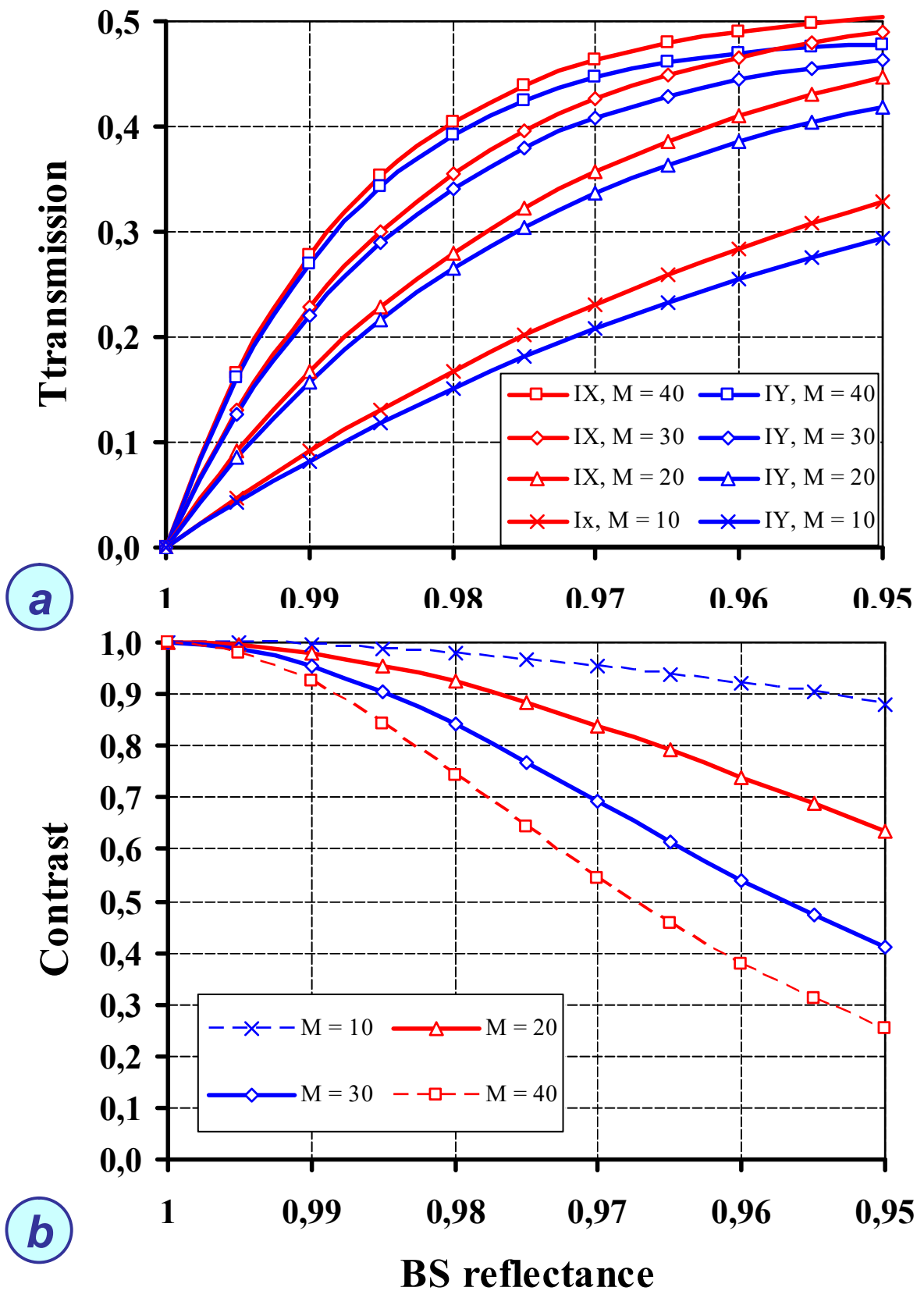

Fig.5: Plots of the transmission curves and of the contrast function of the CSD as functions of the BS reflectance factor for different values of the cavity loop number $M$.

Keeping this in mind and setting all the other parameters to the values indicated in Table 1, the resolving power of the CSD as defined by Eq. 9 is found to be about $1.6 \times 10^{+7}$, an impressive number that is typically one hundred times higher than that of a standard echelle grating, an ten times higher than for a VIPA. However the relative FSR of the CSD is very narrow and can be evaluated to $2.5 \times 10^{-5}$ from Eq. 6. Thus it leads to a moderate number of spectral samples $N_\lambda$ per each diffraction order, varying around 40 in the considered design case. The very high resolving power $R_\lambda$ and the limited FSR of this CSD are illustrated in Figure6, showing a pseudo-color bi-dimensional map of the intensities formed from its exit port +X, as function of the wavelength $\lambda$ and the diffraction angle $\beta$. The sub-picometer resolution level is clearly visible along each horizontal line.

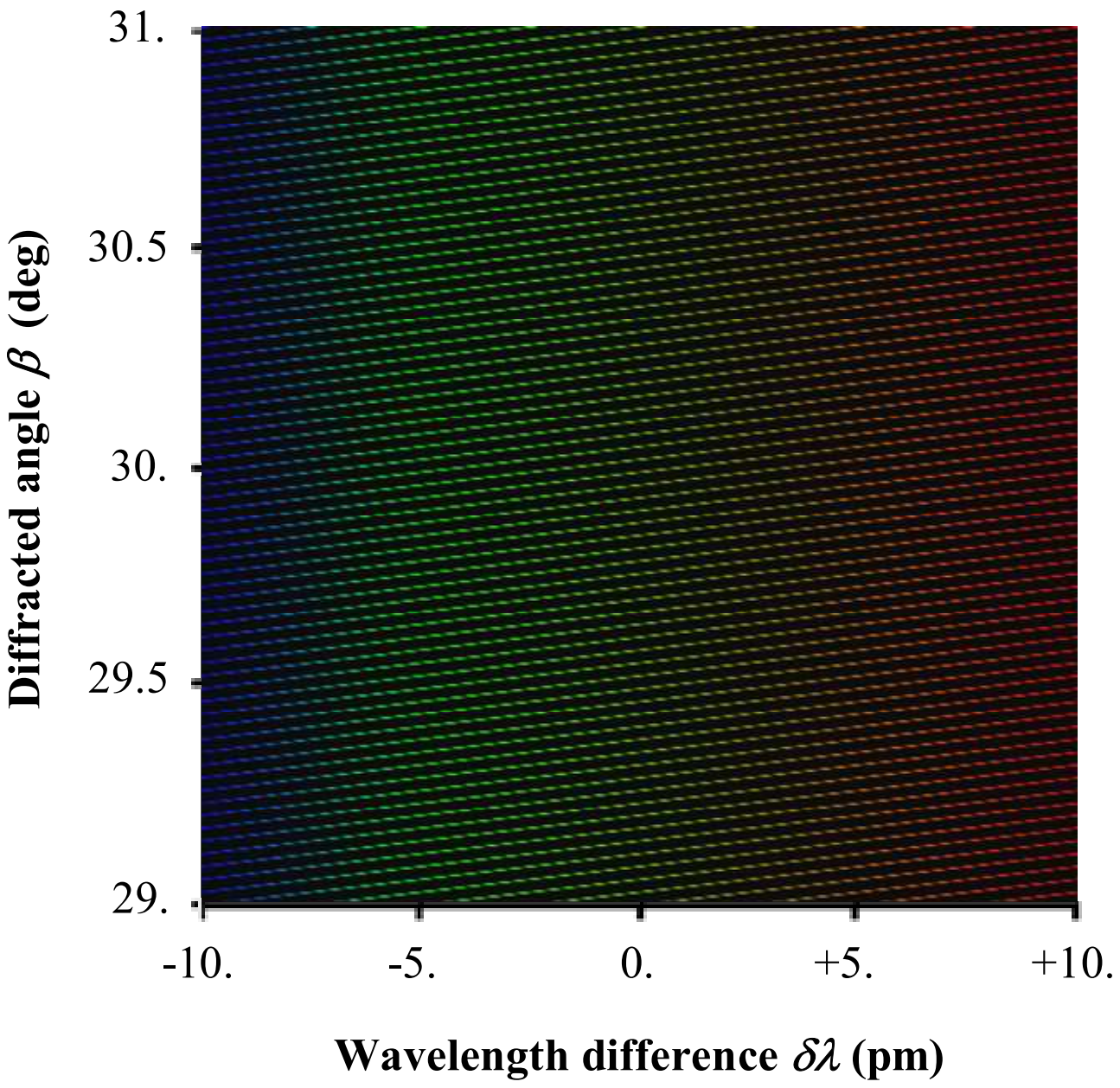

Fig.6: Bi-dimensional map ($\lambda$,$\beta$) of the intensities formed from the exit port +X of the CSD. The pseudo-color scale indicates the lower wavelengths in blue and the higher ones in red. The sub-picometer resolution is clearly visible along each horizontal line.

Table 2: Main characteristics of the proposed CSD design.

| Parameter | Symbol | Value |
|---|---|---|
| Angular dispersion | $\partial\beta / \partial\lambda$ | 1089 rad/μm |
| Spectral resolving power | $R_\lambda$ | 1.59 × 10+7 |
| Free spectral range | $\Delta\lambda / \lambda$ | 2.51 × 10-6 |
| Inversed free spectral range | $\lambda / \Delta\lambda$ | 3.99 × 10+5 |
| Samples number per order | $N_\lambda$ | 40 |
| Transmission on each arm | $I_X$, $I_Y$ | ~ 0.45 |
| Peaks contrast | $C_X$ | 0.7 |

It must be noted that there exist some ways for increasing the spectral resolution of the CSD: for example the hollow cavity formed by the BS and FM depicted in Figure 1 could be replaced with a glass cavity, therefore improving the spectral resolving power proportionally to the refractive index of the chosen glass. Secondly, the previous $L$-side square cavity may be transformed into a rectangular cavity of dimensions $L_X$ x $L_X$ along the X and Y axes respectively, if these lengths can be extended along one or two directions.

## 4. CONCUSION

In this paper was presented the optical design of a long Cavity spectral disperser (CSD) and a theoretical analysis showing that spectral resolving powers about $10^{+7}$ or more are achievable. It may be reasonably hoped that the use of such component may improve significantly the performance of some of the aforementioned techniques, e.g. in the fields of laser shaping, communications and astrophysics, or even detection of gravitational waves.
In the specific context of astronomy, the device may significantly improve our current knowledge of extra-solar planetary science, and especially about rocky planets orbiting in the habitable zone of their parent stars. Here the extreme spectral resolution should be balanced against the flux of photons incoming from the observed extra-solar planet. It is known however that the Signal-to-Nose Ratio (SNR) of the planet is proportional to the square root of the number and strength of hundreds or thousands planet lines detected into the H and K spectral bands by use of cross-correlation techniques for different molecular species (see the review paper from Rukdee, 2024, and the cited references therein). It is also known that this advantage can be enhanced by a few orders of magnitude when using a high-contrast coronagraph located upstream the CSD. For ground-based observatories, it should also allow for better characterization of telluric absorption lines. The device may be used either for planets transiting in front of their parent star or in direct imaging. Both the light transmitted or reflected by the planetary atmospheres can be sensed in the infrared and visible wavebands respectively. Obviously the major science goal of the CSD would be to identifying biological gas markers such as water, methane or oxygen in order to assess the potential habitability of the exoplanet, but it should also give access to the vertical profiles of the partial pressure and temperature of the gases. Moreover, the Doppler shifts of the spectral lines may be used to measure wind speeds in the atmosphere of the planet. Also, extreme spectral resolution allows for mitigating and possibly characterizing the effects of haze and clouds at the planet surface. Taken together, all these features could give rise to a new meteorological and climatic science of extraterrestrial planets.
Finally, the long cavity component described and discussed in this paper may also pave the way to future unexpected and serendipitous discoveries.

**Data Availability**
Data is available on request.